\documentclass[twocolumn,showpacs,twoside,aps, prapp, 10pt, longbibliography, superscriptaddress]{revtex4-1}
\usepackage{graphicx}
\usepackage{epsfig}
\usepackage{epsf}
\usepackage{amssymb}
\usepackage{amsmath}
\usepackage{amsthm}
\usepackage{multirow}
\usepackage{mathrsfs}

\usepackage{graphicx}
\usepackage{dcolumn}
\usepackage{bm}
\usepackage{graphicx}
\usepackage{epsfig}
\usepackage{epsf}
\usepackage{amssymb}
\usepackage{amsmath,empheq,cases}
\usepackage{amsthm}
\usepackage{mathtools}
\usepackage{multirow}
\usepackage[colorlinks=true,linkcolor=blue,citecolor=blue,pdfauthor={ },pdftitle={ },pdfsubject={ },pdfkeywords={ }]{hyperref}
\usepackage{pgfplots}
\usepackage{lipsum}
\usepackage{times}
\definecolor{dred}{rgb}{.8,0.2,.2}
\definecolor{ddred}{rgb}{.8,0.5,.5}
\definecolor{dblue}{rgb}{.2,0.2,.8}
\definecolor{dgreen}{rgb}{.2,0.5,.2}

\newcommand{\RNum}[1]{\uppercase\expandafter{\romannumeral #1\relax}}

\begin{document}

\title{Pulse-width-induced polarization enhancement of optically-pumped N-V electron spin in diamond.}

\author{Yumeng Song}
\thanks{These two authors contributed equally}
\affiliation{School of Electronic Science and Applied Physics, Hefei University of Technology, Hefei, Anhui 230009, China}
\author{Yu Tian}
\thanks{These two authors contributed equally}
\affiliation{Shenzhen Institute for Quantum Science and Engineering and Department of Physics, Southern University of Science and Technology, Shenzhen 518055, China}

\author{Zhiyi Hu}
\affiliation{School of Electronic Science and Applied Physics, Hefei University of Technology, Hefei, Anhui 230009, China}

\author{Feifei Zhou}
\affiliation{School of Electronic Science and Applied Physics, Hefei University of Technology, Hefei, Anhui 230009, China}

\author{Tengteng Xing}
\affiliation{School of Electronic Science and Applied Physics, Hefei University of Technology, Hefei, Anhui 230009, China}

\author {Dawei Lu}
\affiliation{Shenzhen Institute for Quantum Science and Engineering and Department of Physics, Southern University of Science and Technology, Shenzhen 518055, China}

\author{Bing Chen}
\email{bingchenphysics@hfut.edu.cn}
\affiliation{School of Electronic Science and Applied Physics, Hefei University of Technology, Hefei, Anhui 230009, China}

\author{Ya Wang}
\affiliation{Hefei National Laboratory for Physical Sciences at the Microscale and Department of Modern Physics,
University of Science and Technology of China, Hefei 230026, China}
\affiliation{CAS Key Laboratory of Microscale Magnetic Resonance, USTC}
\affiliation{Synergetic Innovation Center of Quantum Information and Quantum Physics, USTC}

\author{Nanyang Xu}
\email{nyxu@hfut.edu.cn}
\affiliation{School of Electronic Science and Applied Physics, Hefei University of Technology, Hefei, Anhui 230009, China}

\author{Jiangfeng Du}
\email{djf@ustc.edu.cn}
\affiliation{Hefei National Laboratory for Physical Sciences at the Microscale and Department of Modern Physics,
University of Science and Technology of China, Hefei 230026, China}
\affiliation{CAS Key Laboratory of Microscale Magnetic Resonance, USTC}
\affiliation{Synergetic Innovation Center of Quantum Information and Quantum Physics, USTC}
\date{\today}

\begin{abstract}

The nitrogen-vacancy (N-V) center in diamond is a widely-used platform for quantum information processing and metrology. The electron-spin state of N-V center could be initialized and readout optically, and manipulated by resonate microwave fields. In this work, we analyze the dependence of electron-spin initialization on widths of laser pulses. We build a numerical model to simulate this process and verify the simulation results in experiment. Both simulations and experiments reveal a fact that shorter laser pulses are helpful to the electron-spin polarization. We therefore propose to use extremely-short laser pulses for electron-spin initialization. In this new scheme, the spin-state contrast could be improved about 10\% in experiment by using laser pulses as short as 4 ns in width. Furthermore, we provide a mechanism to explain this effect which is due to the occupation time in the meta-stable spin-singlet states of N-V center. Our new scheme is applicable in a broad range of NV-based applications in the future.

\end{abstract}

\maketitle

\begin{figure}[h]
\centering
\includegraphics[width=1\columnwidth]{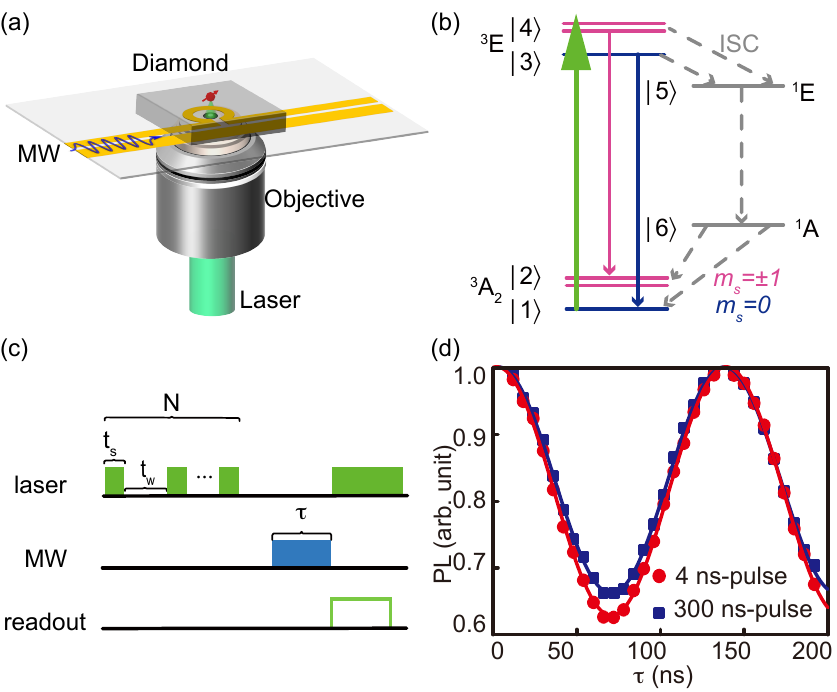} \caption{ Experimental scheme and main result. (a) Experimental setup and a single N-V center in diamond.  (b) Electron-spin energy-level of N-V center and the spin dynamics of the pumping process at room temperature. The transition of pumping laser (532 nm) is indicated by the green solid line, while the radiative (non-radiative) transition in red solid (grey dashed) line. (c) Pulse sequence of the electron-spin Rabi oscillation using repeatedly pulse-width-modulated laser. (d) Effect of pulse-width modulation in electron-spin Rabi oscillation. Blue (red) points are experimental data with $300$ ns ($4$ ns) laser pulses with the pulse sequence shown in (c). Each data point is obtained from $10^{10}$ Rabi sequence repetitions for signal accumulation.}\label{fig1}
\end{figure}

\section{introduction}

N-V center in diamond is a widely-used physical platform for quantum information science and technology \cite{lukin_memory, lukin_science_2006, Ladd2010, Liu2018,  Robledo2011Nature} due to its good controllability and long coherence at room temperature. The electron spin together with nearby nuclear spins works as a quantum register for quantum computation \cite{lukin_science_2006,lukin_science_2007,nv_3bit_entangle,nv_qec,hanson_qec,hanson_weak_nuclear_spin,kf_prl_2016,nv_aqc}, or a nano-scale probe for sensing of magnetic fields \cite{Joreg_ent_sensing,Joerg_broadband,fador_qec_sensing,Joerg_sensing_cluster,Degen_npj_2017,Joerg_chemical_shift,Li2018,Degen2008,Maletinsky2012,Rondin2014,Grinolds2013,Balasubramanian2008,Wang2013,Gruber1997,Maze2008}, electric fields \cite{nv_electric_sensing} and temperatures \cite{Neumann2013,Kucsko2013,Toyli2013}. This also enables further applications of N-V center in  nano-scale nuclear magnetic resonance (NMR) \cite{Joerg_NMR2013,Rugar_NMR_2013,Joerg_chemical_shift, Lukin_NMR} and electron spin resonance (ESR) \cite{Du_single_esr,Bourgeois2015}. A fundamental requirement of these applications is that the spin state of the N-V center can be initialized and read out easily, which is traditionally realized through the same optically pumping process due to a spin-dependent inter-system crossing (ISC) mechanism.

The energy levels of N-V center are shown in Fig. \ref{fig1}b. When pumped with a laser pulse, the electron spin is excited from the  ground spin-triplet ($^3$A$_2$) states to the excited states ($^3$E) \cite{Acosta2010, suter_npol, hanson_optical, Happer1972, nv_n_pol,Lenef1996, nv_electron_dyn}. After a few nanoseconds, the electron spin falls back to the ground state and emits a fluorescence photon. Alternatively, an ISC may happen via meta-stable spin-singlet states ($^1$E and $^1$A) with no fluorescence photon emitted, where the probability from $m_s=\pm1$ to $m_s=0$  is much larger than that of the reverse process \cite{Gaebel2006, Awschalom_polarization, Harrison2006, Gruber1997}. This induces a spin-depended fluorescence photon counts to be detected, hence the spin-state of the N-V center can be read out in experiment. After repeating the process for a few times, the electron spin is prepared in the ground $m_s=0$ with a high probability \cite{Degen2017} . The mechanism of ISC has been studied and attributed to the spin-orbital and electron-phonon interactions \cite{nv_electron_dyn}. Recently, a microscopic model is also provided to explain the electronic states of the N-V center, the optical transitions and the ISC \cite{Goldman2015PRB}.

As a fundamental technique for initialization and readout, optical pumping process has been parametrically studied and the transition rates between the electron-spin states have been measured experimentally \cite{Goldman2015PRL, Goldman2015PRB, nv_electron_dyn}. In a typical N-V experiment at room temperature, the spin polarization of the N-V center to $m_s=0$ is estimated to be  90\% \cite{Hanson_nature_2012}, and a fluorescence contrast between the $m_s=0$ and $m_s=\pm1$ to be over 30\% can be observed when driving the electron spin with a resonant microwave field (\emph{i.e.}, by the Rabi experiment). Furthermore, several new approaches have been proposed to improve the optical readout efficiency of spin states and the effect on nuclear spin polarization during the pumping process is also evaluated in experiments\cite{Joerg_prb_2010, Lukin_PRL_2015, Bassett_prb_2016, Brandt_PRL_2017, Lukin_electron_repetitive_readout, BG_Purcell, Chopped_DNP} .

In the letter, we parametrically study the effect on electron-spin polarizations by using laser pulses with different widths. We change the experimental pumping sequence by repeatedly applying short laser pulses instead of a single long pulse until a steady spin-state is achieved. Numerical simulation shows a increasing polarization as the pulse width decreasing to zero. This result is then verified by observing the spin-state contrasts of Rabi oscillations in experiment. We show that the mechanism to this effect is due to the overall dwelling time of the N-V center in the meta-stale spin-singlet states. Consequently, by reducing the pulse width to 4 ns, we observe an 10\% enhancement in spin-state contrast, which implies an improvement at about 30\% in terms of experimental efficiency. In addition, we evaluate the influences from other experimental parameters, including the laser power, microwave power, and the waiting time between the laser pulses. This work provides a new insight and improvement of efficiency to the optical polarizing process in N-V experiments and is thus useful in future applications.

\section{Results}

\begin{figure}[t]
\centering
 \includegraphics[width=1\columnwidth]{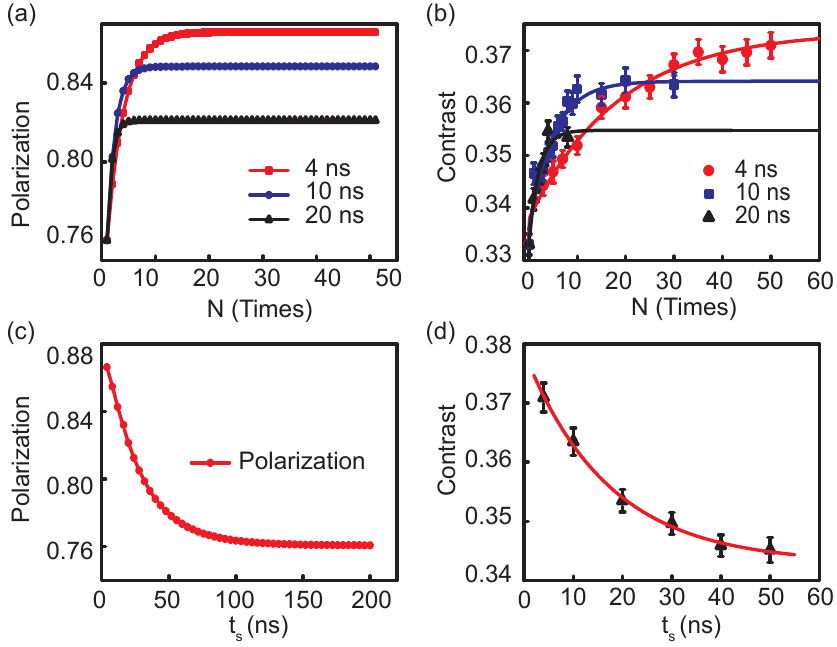}
 \caption{\label{fig2}Numerical simulations and their corresponding experimental results. (a) For laser pulses with three different widths, the highest polarization that can be achieved in simulation is dependent with the repeating times $N$. (b) The measured Rabi oscillation contrast for laser pulses with three different widths and different repetitions. (c) There is a continuous decrease of the highest polarization when the pulse width $t_s$ is increasing from 4 ns to 200 ns in simulation.  (d) The measured polarization (contrast) for pulse widths from 4 ns to 50 ns. Each experimental point is obtained from $10^{10}$ Rabi sequence (Fig. \ref{fig1}c) repetitions for signal accumulation.
 }
 \end{figure}

The essential idea of this work is to examine the dependence of the spin-state polarization on laser pulse widths in the pumping process. The traditional pulse sequence for laser polarization consists of a single square laser pulse followed with a wait time. In this work, we extend this single-loop mode to a more generalized mode as shown in Fig. \ref{fig1}c -- a multi-loop sequence by repeatedly applying laser pulses with a wait time in-between for $N$ times. For each case, $N$ is chosen large enough to saturate the polarization.

In order to evaluate the efficiency of this new sequence, we build a numerical model to simulate the pumping process and calculate the expected polarization in the end of the process. The details of the numerical model and simulation parameters are discussed in the following sections and appendix. Moreover, we apply this new sequence in experiment and parametrically study the performance of the process. Since direct detection of the electron-spin polarization is quite complicated in experiment, we measure the spin-state contrast instead. The contrast is defined as $(I_{max}-I_{min})/I_{max}$,  where $I_{max}$ and $I_{min}$ are the maximal and minimal fluorescence counts fitted from the detected signal in a Rabi experiment, respectively. The whole pulse sequence is shown in Fig. \ref{fig1}c. After polarizing the electron spin, a resonant microwave is applied to drive the spin to oscillate between $m_s=0$ and $m_s=\pm1$ states. Since the spin-state $m_s=0$ $(\pm1)$ is associated with a higher (lower) detected fluorescence counts, the contrast is actually linear with the polarization rate in experiment.

We first examine the pumping process by using different widths of laser pulses in the above sequence. By repeating the pulse loops, we observe the polarization directly in numerical simulations. In Fig. \ref{fig2}a, the results show that, for sequence with shorter laser pulses, the polarization improves more slowly but finally achieves a higher level, and \emph{vice versa}. To verify these trends, we perform the corresponding experiments using the same pulse sequence. For each pulse width, we increase the loop number $N$ and measure the Rabi oscillation signals. The fitted contrast is shown in Fig. \ref{fig2}b, which matches well with the simulated results. Both of the results show that shorter laser pulses are more efficient in the laser pumping process for initialization of the electron spin.

Further, we study the saturated polarization level for each pulse width $t_s$, \emph{i.e.}, the stable value when $N$ is large enough. The simulated result is shown in Fig. \ref{fig2}c, where an exponentially decaying polarization is observed as the pulse width $t_s$ increases. This simulation matches well with the above individual observations. We also verify this numerical expectation with experiments. For each $t_s$, we increase the repeat time $N$ and wait time $t_w$ until the spin-state contrast saturates. The final contrast for each $t_s$ in Fig. \ref{fig2}d matches well with the numerical simulations. Based on these results, we uncover the relation between polarization and pulse width $t_s$.

In experiment, we demonstrate that this new pumping method with repeatedly-applied short laser pulses indeed enhances the polarization in N-V center. For comparison, we measure the Rabi signals using 300 ns (a typical width in traditional experiments) and 4 ns laser pulses using this new method as shown in Fig. \ref{fig1}d, respectively. The experimental result presents a nearly 10\% enhancement in spin-state contrast using this new approach compared to the traditional pumping method. Since the readout stage and the detection noise remain unchanged, this improvement implies around a 30\% save of time cost to achieve the same signal-to-noise ratio (SNR) in future experiments.

\begin{figure}[t]
 \centering
 \includegraphics[width=1\columnwidth]{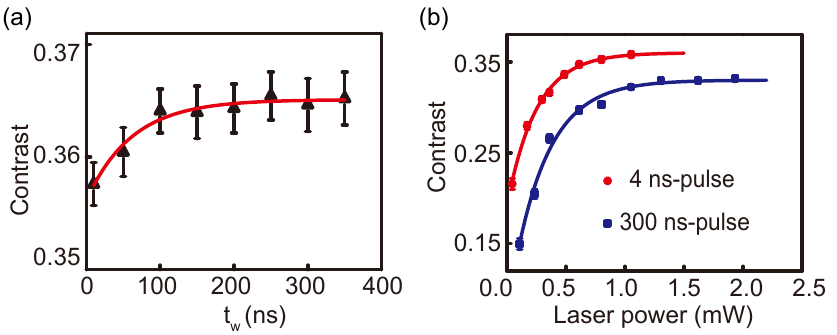}
 \caption{ \label{fig3} Measured Rabi oscillation in terms of spin contrast for (a) different waiting time $t_w$ and (b) different laser powers. Each data point is obtained from $10^{10}$ Rabi sequence (Fig. \ref{fig1}c) repetitions for signal accumulation.
 }
 \end{figure}

To show the stability of our method, we also investigate the influence of different waiting time $t_w$ and laser power in experiment. In Fig. \ref{fig3}a, the waiting time $t_w$ is swept from 10 to 350 ns and the contrast is measured accordingly, where we set $t_s = 4$ ns and $N = 30$. The experimental result shows that, for $t_w$ greater than 100 ns, the polarization efficiency remains almost stable. For the laser powers, we test both the traditional single-loop (the 300 ns pulse) and our multi-loop schemes (the 4 ns pulse) in Fig. \ref{fig3}b. In both cases, the polarization performance is proportional to the laser power, and saturates when the power is over 1 mW. Hence, it can be concluded that for either the traditional scheme or our scheme, the effect of the laser power is almost irrelevant in boosting the polarization performance, except for a slight discrepancy between the 4 ns and the 300 ns curve. This is mainly attributed to the low efficiency of the acousto-optic modulator (AOM) working in the short-pulse regime.

\section{Numerical Simulations}

Here we describe the process for numerical simulations. Let us refer to the schematics of the electron-spin energy levels and laser pumping process in Fig. \ref{fig1}b.  The ground (excited) state is a spin triplet $^3$A$_2$ ($^3$E) with energy splitting $D_{gs} = 2.87$  GHz ($D_{es} = 1.41 $ GHz) between $m_s=0$ and $m_s=\pm1$ states at zero magnetic field.  Since the dynamics in $m_s=+1$ and $-1$ are equivalent in the pumping process, we use a reduced six-level system without loss of generality, including the ground states ($\left| 1 \right\rangle$, $\left| 2 \right\rangle$), excited states ($\left| 3 \right\rangle$, $\left| 4 \right\rangle$), and the spin-singlet system involved in ISC ($\left| 5 \right\rangle$, $\left| 6 \right\rangle$). We employ a rate equation model to simulate the transition process. Here, the state of the N-V center is stored in a vector  $\textbf{P} = (P_1,P_2,P_3,P_4,P_5,P_6)$ with $P_i$ the population of state $\left| i \right\rangle$ and $\sum_{i}{P_i}=1$. The population vector changes during the pumping process according to equations which can be expressed in the matrix form as:
\begin{equation}
\frac{d}{dt}\textbf{P}= M_i\textbf{P},
\label{eq1}
\end{equation}
where $M_{0} (M_{1})$ denotes the transition matrix when laser illumination is on (off) and the values of the matrix elements are shown in the appendix.

\begin{figure}[t]
\centering
\includegraphics[width=1\columnwidth]{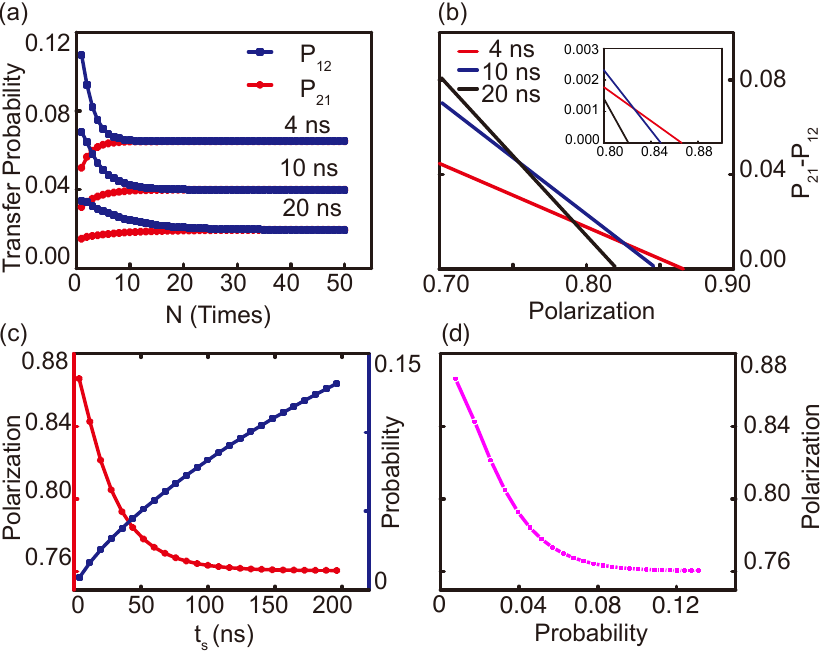}
\caption{  (a) In simulation, the population transfer $P_{21}$ and $P_{12}$ eventually converge to the same value as the repetition $N$ increases.  Here, $P_{ij}$ denotes the population transfer from $\left|i\right\rangle$ to $\left|j\right\rangle$. (b) Difference between $P_{21}$ and $P_{12}$, which indicates the net transfer between $\left|1\right\rangle$ and $\left|2\right\rangle$. (c) Probability (blue) at meta-stable spin-singlet states ($|5\rangle$ and $|6\rangle$)  and the corresponding final polarization (red) with different $t_s$. (d) Direct relation between the final polarization and the probability at meta-stable spin-singlet states.}\label{fig4}
 \end{figure}

The pumping process in a typical N-V experiment consists of a single square laser pulse and a wait time as shown in Fig. \ref{fig1}c. The target of this process is to initialize the electronic spin into the $m_s = 0$ ground state with high probability. Here, in order to analyze the pulse-width-induced effect on spin polarization, we replace the traditional single square laser pulse with $N$ repeatedly-applied short pulses, $i.e.$, the setting in our scheme. As shown in Fig. \ref{fig1}d, each short pulse is of width $t_s$ and an wait time $t_w$ is inserted in-between. We simulate this process using the above equation and calculate the polarization in the end of this process. In simulations with fixed parameters $t_s$ and $t_w$, the polarization is approaching a stable level when the repeating times $N$ increases as in Fig. \ref{fig4}a. Thus to get the best polarization for each $t_s$ or $t_w$, we choose $N$ to be large enough (\emph{e. g.}, 400 for most of the cases). 

\section{The mechanism}

Here, we consider the case where the pumping process starts with a thermal state $\textbf{P}=(\frac{1}{3}, \frac{2}{3},0,0,0,0)$. For a single pulse loop with a short laser pulse and a wait time, the increased polarization depends on the value $P_{21}-P_{12} > 0$, where $P_{ij}$ denotes the transferred population from state $\left| i \right\rangle$ to $\left| j \right\rangle$ in a single loop. As the pulse loop starts to repeat, the population $P_2$ increases while $P_1$ decreases, thus $P_{21}$ goes smaller and $P_{12}$ goes bigger. Finally, a steady state is obtained with $P_{12}=P_{21}$ and the pumping process is completed. This process is show in Fig. \ref{fig4}a, where for different $t_s$ the value $P_{21} (P_{21})$ changes with different speeds. This difference induces different final polarizations of the steady states. We also show this effect by plotting the the net transferred population $P_{21}-P_{12}$ in a single loop towards the polarization in Fig. \ref{fig4}b. For each pulse width $t_s$, the best polarization is achieved in the steady state where $P_{21}-P_{12} = 0$. From this simulation, we see that smaller $t_s$ is associated with better polarizations, which matches well with experimental results.

The reason to this pulse-width induced polarization difference, however, is due to the overall population of the N-V center in the spin-singlet meta-stable states, \emph{i.e.}, state $\left| 5 \right\rangle$ and $\left| 6 \right\rangle$. During the initialization process, the transferred population from state $\left| 3\right\rangle$ ($\left| 4 \right\rangle$) to $\left| 5 \right\rangle$ increases (decreases) until a steady state is reached. Once the N-V center is on the state $\left| 5 \right\rangle$, It falls down to state $\left| 6 \right\rangle$ immediately, and subsequently onto the ground state $\left| 1 \right\rangle$ or $\left| 2 \right\rangle$. Since the transition rates from state $\left| 6 \right\rangle$ to state $\left| 1 \right\rangle$ and  $\left| 2 \right\rangle$ are almost the same, this transition transfers similar amounts of population to the ground spin-states and reduce the final polarization of the whole pumping process.

Therefore, reducing the dwelling time on the meta-stable states would increase the final polarization level in the process. To confirm, we integrate the populations on $\left| 5 \right\rangle$ and $\left| 6 \right\rangle$  during the pumping process (including the wait time) in numerical simulations. The relation between the integrated population and laser pulse widths is shown in Fig. \ref{fig4}c. We see that, the integrated population decreases as the pulse width decreases. In addition, we show the direct relation between the final polarization and the integrated population in Fig. \ref{fig4}d, where a close-to-linear relation is observed. These results can be used to explain the mechanism of the pulse-width induced polarization enhancement, which is in fact due to the reduction of the dwelling time on the meta-stable spin-singlet states during the pumping process.

\section{Experiments}

To demonstrate, we use a home-built optically-detected magnetic resonance (ODMR) system to address and manipulate the single NV-based centers in a type-\RNum{2}a, single-crystal synthetic diamond sample (Element Six). As shown in Fig. \ref{fig1}a, the 532 nm pumping laser is focused on the sample via a 100X oil-immersed objective. The emitted fluorescence ranging from 650 to 800 nm is spatially filtered through a 50 $\mu$m pinhole and finally collected by an avalanche photodiode (APD) for single-photon counting. A low magnetic field with a few Gauss is applied along the N-V axis to split the energy levels of $m_s=\pm1$. A microwave field  used to drive the electron-spin state is generated from Microwave Source (Rohde\&Schwarz SMIQ03) and amplified by a wide-band amplifier (Mini-circuits ZHL-42W) to drive the transition between $m_s=0$ and $m_s=-1$. An impedance-matched copper slot line with gap of 0.1 mm with an $\Omega$-type ring (inner diameter 300 $\mu$m) mounted on a coverslip is used to hold the sample and radiate the microwave fields.

In this work, the pumping laser beam is modulated by a 350-MHz AOM before applied on the sample. The AOM is driven by a 2.6-GSPS AWG (Tektronix AWG610) with an output bandwidth over 800 MHz, which is capable of generating laser pulses as short as 4 ns. In order to suppress the laser leakage, another AOM is used following the first one, which turns off the laser in the rest time of experiment. In order to measure the spin-state contrast, we perform Rabi oscillation experiment shown in Fig. \ref{fig1}c , where the initialization with traditional square (repeatedly shorter) laser pulses is applied on the N-V center followed with a resonant variable-width microwave pulse and another laser pulse for state readout. The signal is fitted with a cosine function, and the contrast is calculated with the fitting parameters.

\section{conclusion}

To conclude, we analyze the pulse-width-induced effect on the electron-spin polarization of the N-V center in the optical initialization process, and provide a new scheme to polarize the electron spin with repeatedly-applied short laser pulses. This new scheme provides an enhancement about 10\% in readout efficiency, leading to an about 30\% save of time cost in experiment. Moreover, we build a numerical model to simulate the laser initialization process and calculate the dependence of the polarization under different parameters. The result matches well with the experimental observations of spin-state contrast. By analyzing the mechanism of our new scheme, we conclude that the superiority of our method is mainly due to the reduction of the dwelling time on the meta-stable spin-singlet states during the initialization process. Our new scheme could be applied to the NV-based quantum applications in a broad range, and may shed light in understanding the optical initialization process in the N-V centers.

\begin{acknowledgments}
This work is supported by the National Key R\&D Program of China (Grants No. 2018YFA0306600, No. 2017YFA0305000 and No. 2019YFA0308100 ), the Fundamental Research Funds for the Central Universities (Grants No. PA2019GDQT0023), the NNSFC (No. 11761131011,No. 11775209 and No. 11875159),  the CAS (Grants No. GJJSTD20170001 and No. QYZDY-SSW-SLH004), Anhui Initiative in Quantum Information Technologies (Grant No. AHY050000) and the Innovative Program of Development Foundation of Hefei Center for Physical Science and Technology (Grants No. 2017FXCX005).

\appendix
\section{Simulation Parameters}

In the pumping process, the transition rate is defined as  $k_{ij}$ which refers to the transfer speed from state $\left| i \right\rangle$ and $\left| j \right\rangle$. The detailed parameters of $k_{ij}$ are listed in Tab.\ref{tab:table1}. In the numerical simulation, the state of electrons is calculated from the equation in each piece of time which can be expressed in the matrix form as $\frac{d}{dt}\textbf{P}= M_i\textbf{P}$. The matrix $M_{0}(M_{1})$ is defined as
\begin{widetext}

\begin{equation}
M_0 =\left[
    \begin{array}{cccccc}
        -k_{13}&0&k_{31}&k_{41}&0&k_{61}\\
        0&-k_{24}&k_{32} & k_{42} & 0 & k_{62}\\
        k_{13} & 0 & -(k_{32}+k_{31}+k_{35}) & 0 & 0 &0\\
        0& k_{24}& 0 & -(k_{42}+k_{41}+k_{45}) & 0 & 0\\
        0& 0& k_{35} & k_{45} & -k_{56} & 0\\
        0& 0& 0& 0& k_{56}& -(k_{61}+k_{62})
    \end{array}
\right]\label{eq2}
\end{equation}
and
\begin{equation}
 M_1 =\left[\begin{array}{cccccc}
        0&0&k_{31}& k_{41}&0&k_{61}\\
        0&0&k_{32}& k_{42} & 0 & k_{62}\\
        0& 0 & -(k_{32}+k_{31}+k_{35})&0 & 0 &0\\
        0&0 & 0& -(k_{42}+k_{41}+k_{45}) & 0 & 0\\
        0& 0& k_{35}& k_{45} & -k_{56} & 0\\
        0& 0& 0&0& k_{56}& -(k_{61}+k_{62})
        \end{array}\right].\label{eq3}
\end{equation}

\begin{table}[h]
\caption{\label{tab:table1}  Transition Rates }
\begin{ruledtabular}
 \begin{tabular}{cccc}
  $k_{ij}$&Transition Rate(GHz)&$k_{ij}$&Transition Rate(GHz)\\
  \hline
  $k_{13}$,$k_{24}$ & 0.628 & $k_{45}$ & 0.1884\\
  $k_{31}$,$k_{42}$ & 0.4396 & $k_{56}$ & 6.28\\
  $k_{32}$,$k_{41}$ & 0 & $k_{61}$ & 0.020724\\
  $k_{35}$ & 0.0314 & $k_{62}$ & 0.013816\\
  \end{tabular}
 \end{ruledtabular}
\end{table}
\end{widetext}

\end{acknowledgments}

%

\end{document}